\documentclass[11pt]{article}

\usepackage{a4wide,graphics,graphicx,amsmath,amssymb,cite,nicefrac,upgreek}
\usepackage[verbose]{wrapfig}
\usepackage{physics}
\usepackage{authblk,hyperref}
\hypersetup{
	colorlinks=true,
	linkcolor=blue,
	filecolor=red,      
	urlcolor=cyan,
	citecolor=red
} 
\usepackage{tikz}
    \usepackage{pgfplots}
    \pgfplotsset{compat=1.17} 
\usepgfplotslibrary{fillbetween}
\usetikzlibrary{positioning,decorations.pathmorphing}
\catcode`@=11 \@addtoreset{equation}{section} \catcode`@=12

\begin{document}
\date{}
\title{
{\baselineskip -.00000000005in
} 
\vskip .000000000000000000002cm
\vbox{
{\bf \Large Noncommutative black holes: Topological bulk–boundary correspondence and Binary Merger Bounds}
}}

\author{Ankit Anand\thanks{email: anand@iitk.ac.in}}
\affil{\normalsize\it Department of Physics, Indian Institute of Technology Kanpur, Kanpur 208016, India}

\author{Anshul Mishra\thanks{email: anshulmishra2025@gmail.com}}
\affil{\normalsize\it Department of Physics, RRSDCE, Begusarai - 851134, India.}

\author{Aditya Singh\thanks{email: 24pr0148@iitism.ac.in}}
\affil{\normalsize\it Department of Physics, Indian Institute of Technology (Indian School of Mines) Dhanbad, Jharkhand 826004, India}

\author{Saeed Noori Gashti\thanks{email: saeed.noorigashti@stu.umz.ac.ir; sn.gashti@du.ac.ir}}
\affil{\normalsize\it School of Physics, Damghan University, P. O. Box 36716-41167, Damghan, Iran.}

\author{Neeraj Kumar\thanks{email: nkneeraj06@gmail.com}}
\affil{\normalsize\it School of Science, Walailak University, Nakhon Si Thammarat, 80160, Thailand.}

\author{Phongpichit Channuie\thanks{email: phongpichit.ch@mail.wu.ac.th (Corresponding author)}}
\affil{\normalsize\it College of Graduate Studies, Walailak University, Nakhon Si Thammarat, 80160, Thailand.}

\maketitle

\begin{abstract}
We investigate the thermodynamic topology of charged AdS black holes in a non-commutative spacetime sourced by Lorentzian-smeared matter distributions. Since exact analytical solutions for the critical thermodynamic quantities are not available, we employ a perturbative expansion in the non-commutative parameter and validate the resulting expressions through numerical analysis. Using the generalized off-shell free-energy framework, we explore the topological structure of the thermodynamic phase space and evaluate the corresponding winding number that characterizes the phase transitions. Our results reveal that non-commutative effects introduce qualitative modifications to the thermodynamic behavior compared with the standard Reissner–Nordström AdS black hole. Furthermore, we demonstrate that the bulk and boundary descriptions possess an identical global thermodynamic topology, providing strong evidence for the correspondence between their topological structures. We also investigate the lower bound on the remnant mass implied by the second law of black-hole thermodynamics and observe that non-commutative corrections modify key thermodynamic quantities, with particular emphasis on the entropy and the final black-hole mass.

\end{abstract}

\newpage

\section{Introduction}

Black hole thermodynamics~\cite{Bardeen:1973gs} offers a powerful framework for exploring the interface between gravity, quantum theory, and statistical mechanics. Bekenstein~\cite{Bekenstein:1972tm, Bekenstein:1973ur} proposed that a black hole carries an entropy proportional to the area of its event horizon and firmly established by Hawking’s demonstration that black holes radiate thermally due to quantum field effects in curved spacetime, with temperature $T_H=\kappa/(2\pi)$, where $\kappa$ denotes the surface gravity~\cite{14}. These insights make black holes to genuine thermodynamic systems and have motivated extensive investigations of their stability, phase structure, and critical behaviour~\cite{Bekenstein:1974ax, Hawking:1975vcx, Gibbons:1976ue, Brown:1994gs, Allen:1984bp, York:1986it, Whiting:1988qr}. Despite significant progress, the final states of black hole evaporation remain poorly understood, largely due to the absence of a complete theory of quantum gravity. 

String theory and loop quantum gravity are widely regarded as two leading candidates for a consistent theory of quantum gravity. Both frameworks predict that classical spacetime geometry must be modified at very short length scales, typically of the order of the Planck length. In string theory, this expectation is supported by the string/black hole correspondence principle, which asserts that as a black hole shrinks toward the string scale, its description smoothly transitions from a classical black hole to an excited string state \cite{Susskind:1993ws, Horowitz:1996nw}. This correspondence implies that purely classical geometrical notions cease to be valid near the Planck regime, and that stringy degrees of freedom provide an effective description of quantum gravitational effects. Similarly, loop quantum gravity predicts a fundamentally discrete structure of spacetime, leading to modifications of classical singularities and horizon properties through quantum geometric corrections \cite{Rovelli:1997qj, Ashtekar:2004eh}. These insights strongly motivate the use of effective models that incorporate short-distance quantum corrections to explore gravitational phenomena beyond the classical regime.

A particularly well-motivated framework for encoding short-distance quantum corrections is provided by noncommutative geometry, in which spacetime coordinates are promoted to noncommuting operators satisfying
\begin{eqnarray}
    [x^\mu,x^\nu] = i\,\Theta^{\mu \nu}\, .
\end{eqnarray}
Here, the antisymmetric tensor $\Theta^{\mu \nu}$ introduces a fundamental length scale that effectively discretizes spacetime, in close analogy with the role of Planck’s constant in phase space. It has been shown that Lorentz invariance and unitarity can be consistently preserved within the Weyl-Wigner-Moyal formulation by adopting a canonical structure for $\Theta^{\mu \nu}$, thereby providing a controlled and physically viable deformation of classical field theories and gravity \cite{Seiberg:1999vs,Chaichian:2001py}. Notably, noncommutative geometry arises naturally in string theory when open strings propagate in the presence of a background antisymmetric $B$-field on D-branes, leading to spacetime coordinates that obey nontrivial commutation relations \cite{Seiberg:1999vs,Douglas:2001ba}. The associated noncommutative parameter thus sets an intrinsic length scale, encoding quantum spacetime effects beyond the classical continuum description.

There are two distinct approaches in the literature for studying the noncommutative-inspired theory. One is based on algebraic deformations of spacetime symmetries, implemented through Drinfel’d twists and the associated Seiberg-Witten map \cite{Drinfeld:1983ky, Seiberg:1999vs}. In this formulation, noncommutativity is encoded in the level of spacetime coordinates and symmetry algebras, leading to deformed field equations and modified geometric structures. At leading order, the spacetime metric generically acquires off-diagonal components governed by model-independent correction functions, reflecting the intrinsic anisotropy introduced by the twist deformation. The other one is a complementary and widely adopted approach that models noncommutativity through the smearing of point-like mass and charge distributions over a minimal length scale, resulting in regularized black hole geometries and controlled perturbative corrections to thermodynamic quantities \cite{Nicolini:2005vd, Ansoldi:2006vg}. This effective description has proven particularly useful for investigating black hole thermodynamics, critical behavior, and supercritical crossover phenomena, including the emergence of Widom lines and scaling properties. These models have been extensively explored in recent years, with applications ranging from black hole thermodynamics and Hawking radiation, quasinormal mode spectra and many others~\cite{30, Gupta:2022oel, Herceg:2023zlk, Herceg:2023pmc, Herceg:2024vwc, Herceg:2025fyf, Herceg:2024upt, Mann:2011mm, Lopez-Dominguez:2006rou, Modesto:2010rv, Wang:2024jlj}

The discovery that black hole thermodynamics can be understood from a topological standpoint has recently enhanced the field~\cite{a19, a20}. Rather than relying primarily on local thermodynamic response functions, topological techniques capture phase transitions using global invariants, such as winding numbers and defect topologies, that are insensitive to microscopic system details. Thermodynamic phases are determined in this paradigm using topologically unique configurations of auxiliary vector fields built from off-shell thermodynamic potentials. Phase transitions and stability features have been effectively captured over a wide class of black hole solutions~\cite{a19, a20, Du:2023nkr, Wu:2022whe, Wu:2023sue, Liu:2022aqt} using methods based on topological current theory and free-energy topology. These developments provide strong motivation to investigate whether noncommutative black holes exhibit similar topological signatures and to assess how quantum spacetime effects influence the global organization of black hole phases. The method has since been extended to a broad class of gravitational solutions \cite{a19, a20, 20a, 21a, Du:2023nkr, Liu:2022aqt, Wu:2023sue, Wu:2022whe, 22a, 23a, 24a, 26a,27a, 35a, 38', Alipour2024, 38a, 38b, 38c,40a,42a, 43a, 44a, 31a,33a,34a, 44g, 44h, 44i, 44j, 44k, 44l, 44m}.

Holographically, the physical interpretation of the cosmological constant \(\Lambda\) in extended black hole thermodynamics remains subtle, since its counterpart in the dual conformal field theory is not uniquely understood~\cite{Visser:2021eqk}. The thermodynamic features of AdS black holes encode properties of a strongly coupled boundary theory at large \(N\)~\cite{Maldacena:1997re, Witten:1998qj}. While \(\Lambda\) plays the role of pressure in the bulk thermodynamic description, this identification does not translate directly to a pressure variable in the boundary theory. Instead, changes in \(\Lambda\) are widely interpreted as modifying the number of degrees of freedom of the dual theory, typically captured by the central charge and controlled by the AdS curvature scale. As the central charge fixes the effective size of the boundary system, varying \(\Lambda\) naturally induces a change in the boundary volume. This perspective has led to a consistent formulation of extended black hole thermodynamics on the boundary, initially within Einstein gravity~\cite{Ahmed:2023snm, Gong:2023ywu, Ahmed:2023dnh, Baruah:2024yzw} and subsequently generalized to more elaborate holographic settings, including nonrelativistic geometries and theories with a finite radial cutoff~\cite{Cong:2024pvs, Zhang:2025dgm}. Several studies have provided evidence for a bulk–boundary correspondence in thermodynamic topology \cite{Zhang:2023uay, Yang:2025uul}. Extending this analysis to noncommutative spacetimes is well motivated, as noncommutativity provides a concrete and controllable framework for implementing UV regularization and probing how quantum geometric effects modify bulk–boundary thermodynamic topology.

In addition to providing insights into equilibrium properties, black-hole thermodynamics also imposes fundamental constraints on dynamical processes such as binary black-hole mergers. The generalized second law requires that the entropy of the post-merger remnant must be at least as large as the sum of the entropies of the initial black holes, thereby establishing a lower bound on the final mass and an upper bound on the energy that can be emitted through gravitational radiation \cite{Bekenstein:1974ax, Hawking:1971vc}. With the advent of gravitational-wave astronomy and the growing catalog of compact-binary coalescence events detected by the LIGO--Virgo--KAGRA collaborations, such thermodynamic arguments have acquired renewed significance as probes of theories beyond general relativity \cite{Abbott:2016blz, Abbott:2020khf}. Since noncommutative geometry introduces a fundamental minimal length scale that modifies the horizon structure and entropy of black holes, it is natural to investigate how these corrections influence the thermodynamic constraints on merger outcomes. In this work, we employ the generalized second law to derive a lower bound on the remnant mass of binary mergers involving noncommutative black holes and examine how the noncommutative parameter alters the allowed post-merger configurations relative to the commutative limit.

The paper is organized as follows: In Sec.~\ref{Sec:Non-Commutative Charged Black hole}, we review the noncommutative extension of the Reissner–Nordström–AdS black hole using Lorentzian-smeared mass and charge distributions and discuss the corresponding metric functions. In Sec.~\ref{Sec:Thermodynamics and criticality with noncommutative parameter}, we develop the perturbative thermodynamic framework, compute the modified thermodynamic quantities, and analyze the critical behavior in the extended phase space. In Sec.~\ref{Sec:Topological Study}, topological study is being done. In Sec.~\ref{Sec:Topological Charge in Dual CFT} we compute the topological charge in dual CFT. Thereafter, we study the  Binary Black Hole Merger in Sec.~\ref{Sec: Binary Black Hole Merger} and finally, Sec.~\ref{Sec:Summary and Discussions} summarizes our results and discusses the implications of noncommutativity for AdS black hole thermodynamics and possible quantum-gravity signatures.


\textbf{Note:} It is important to emphasize that the aim of the present work is not merely to study the critical behavior reported in~\cite{Hadri:2025mvu}, but to revisit the analysis while keeping the noncommutative parameter explicit throughout the thermodynamic treatment. In~\cite{Hadri:2025mvu}, the auxiliary quantity \(b=2\alpha/(3r_+)\) was introduced and held fixed when implementing the criticality conditions. Since the specific volume satisfies \(v\propto r_+\), this procedure effectively relates the deformation parameter \(\alpha\) to the thermodynamic variable with respect to which the derivatives are taken. Consequently, the resulting criticality analysis probes constant-\(b\) trajectories in parameter space rather than configurations with fixed noncommutative coupling. In contrast, we impose the inflection-point conditions at fixed \(\alpha\), treating the deformation parameter as an intrinsic characteristic of the underlying theory. We believe that this prescription provides a more systematic and thermodynamically consistent characterization of the effects of noncommutativity on the critical behavior. While a similar procedure is also used in~\cite{Liu:2016uyd, Belmahi:2025ysk}, such a parametrization may implicitly restrict the set of independent thermodynamic variables and may therefore obscure certain aspects of the phase structure.

\section{Review of Noncommutative Charged Black hole}\label{Sec:Non-Commutative Charged Black hole}
Inspired by the role of Non-commutative (NC) geometry in string theory and quantum gravity, one constructs the NC extension of the Reissner–Nordström–Anti–de Sitter (RN–AdS) black hole. In NC geometry, spacetime coordinates are promoted to noncommuting operators,
\begin{equation}
[x^\mu , x^\nu] = i\Theta^{\mu\nu} \ ,
\end{equation}
where $\Theta^{\mu\nu}$ is a constant antisymmetric tensor. This deformation introduces a minimal length scale $\sqrt{\Theta}$, effectively regularizing point-like singularities. Such corrections, linked to the background $B_{\mu\nu}$ field in string theory \cite{15}, modify the horizon structure, thermodynamics, and black hole stability. We use the idea in which NC effects are captured by replacing the point-like mass and charge distributions by smooth, spherically symmetric ``smeared'' profiles with a characteristic length scale $\sqrt{\Theta}$. 

In this subsection, we review NC RN-AdS black hole solutions sourced by a standard Maxwell field in the presence of a negative cosmological constant, $\Lambda = -\tfrac{3}{\ell^2}$, where $\ell$ denotes the AdS radius. The dynamics are governed by the Einstein-Maxwell action
\begin{equation}
S = \int d^4x \sqrt{-g}\left( R  -\frac{6}{\ell^2} + 2F_{\mu\nu}F^{\mu\nu} \right) \ .
\label{action}
\end{equation}
Varying the action with respect to the metric and the gauge field yields the coupled field equations
\begin{equation}
R_{\mu\nu} - \frac{1}{2}g_{\mu\nu}R -\frac{3}{\ell^2} g_{\mu\nu} = 8\pi \left( T^{\text{matt}}_{\mu\nu} + T^{\text{el}}_{\mu\nu} \right) \qquad;\qquad \nabla_\mu F^{\mu\nu} = J^\nu \ .
\label{field_eqs}
\end{equation}
Within the NC geometry-inspired framework, point-like sources are replaced by smeared matter distributions characterized by a fundamental length scale $\sqrt{\Theta}$. The matter sector is described by an anisotropic fluid with a stress-energy tensor in the diagonal form as  $(-\rho_{\text{matt}}(r,\Theta),\,\rho_{\text{matt}}(r,\Theta),\,p_\theta(r,\Theta),\,p_\phi(r,\Theta)),$ where the angular pressures satisfy
\begin{equation}
p_\theta = p_\phi
= -\rho_{\text{matt}}(r,\Theta)
+ \frac{r}{2}\frac{d\rho_{\text{matt}}(r,\Theta)}{dr}.
\label{pressure_relation}
\end{equation}

The electromagnetic contribution is given by the standard Maxwell stress energy tensor. Assuming a purely electric, spherically symmetric configuration, the field strength tensor and current density are chosen as
\begin{equation}
F_{tr} =  E(r,\Theta) \qquad;\qquad
J^\nu = \rho_{\text{el}}(r,\Theta)\,\delta^\nu_0 .
\label{field_strength}
\end{equation}
Using the Lorentzian smearing functions, which are widely used in NC black hole models. The Lorentzian mass and charge distributions,
\begin{equation}
\rho_{\text{matt}}(r,\Theta) = \frac{M\sqrt{\Theta}}{\pi^{3/2}(r^2+\pi\Theta)^2} \qquad ; \qquad \rho_{\text{el}}(r,\Theta) = \frac{Q\sqrt{\Theta}}{\pi^{3/2}(r^2+\pi\Theta)^2} \ ,
\label{lorentzian_density}
\end{equation}
These distributions reduce to point-like sources in the commutative limit $\Theta \to 0$.

We adopt the static, spherically symmetric line element
\begin{equation}\label{Metric_function}
ds^2 = -f(r)\,dt^2 + \frac{dr^2}{f(r)} + r^2 d\Omega^2 \quad \text{with}\quad f(r)=1-\frac{2\,m(r)}{r}+\frac{q(r)^2}{r^2}+\frac{r^2}{\ell^2} \ .
\end{equation}
Using Eq.~\eqref{field_strength}, Maxwell's equation for the enclosed charge $q(r)$ with the relation $E(r)=q(r)/r^2$ reduces to
\begin{equation}\label{eq:qprime}
\frac{d q(r)}{dr}=4\pi r^2 \rho_{el}(r) \quad \Longrightarrow \quad q(r)=4\pi\int_0^r \rho_{el}(r')\,r'^2\,dr' \ .
\end{equation}
Finally, solving the Einstein's equation using Eq.~\eqref{lorentzian_density}, one obtains a relation between the matter density and the mass function, with the convention that all the mass stored in the
matter distribution we have
\begin{equation}\label{eq:mprime}
m'(r)=4\pi r^2 \rho_M(r) \quad\Longrightarrow\quad m(r)=4\pi\int_0^r \rho_M(r')\,r'^2\,dr' \ .
\end{equation}

Performing the radial integrals in Eq.~\eqref{eq:qprime} and Eq.~\eqref{eq:mprime} with Eq.~\eqref{lorentzian_density} yields the explicit enclosed functions
{\small\begin{align}\label{eq:mr}
m(r) &= \frac{2M}{\sqrt{\pi}}\left[ \arctan\!\left(\frac{r}{\sqrt{\Theta}}\right) - \frac{r\sqrt{\Theta}}{r^2+\Theta} \right]\quad ; \quad q(r) = \frac{2Q}{\sqrt{\pi}}\left[ \arctan\!\left(\frac{r}{\sqrt{\Theta}}\right) - \frac{r\sqrt{\Theta}}{r^2+\Theta} \right] \ . 
\end{align}}

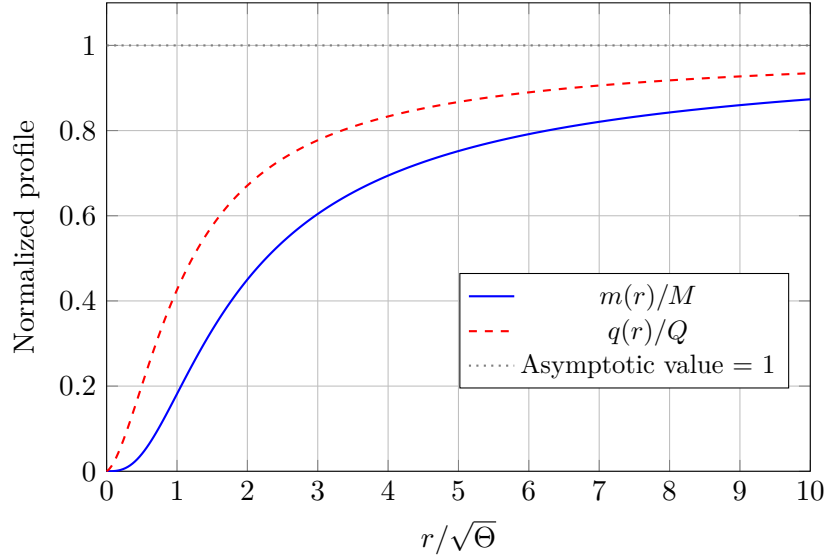
\begin{figure}[ht]
\centering
\begin{tikzpicture}
  \begin{axis}[
    width=0.7\textwidth,
    height=0.5\textwidth,
    xlabel={$r/\sqrt{\Theta}$},
    ylabel={Normalized profile},
    xmin=0, xmax=10,
    ymin=0, ymax=1.1,
    grid=both,
    legend style={at={(0.97,0.3)},anchor=east,font=\small}
  ]
    \addplot[domain=0:10,samples=300,thick,blue]
      {(2/pi)*(rad(atan(x)) - (x)/(x^2+1))}; 
    \addlegendentry{$m(r)/M$};

    \addplot[domain=0:10,samples=300,thick,red,dashed]
      {sqrt((2/pi)*(rad(atan(x)) - (x)/(x^2+1)))}; 
    \addlegendentry{$q(r)/Q$};

    \addplot[gray,dotted,thick,domain=0:10]{1};
    \addlegendentry{Asymptotic value = 1};
  \end{axis}
\end{tikzpicture}
\caption{Normalized mass and charge profiles for Lorentzian smearing with $\Theta=1$.}
\label{fig:mass-charge-profiles}
\end{figure}
By putting this in Eq.~\eqref{Metric_function}, the metric function for the NC-corrected charged black hole is 
{\footnotesize \begin{eqnarray}\label{Working_metric_function1}
    f(r) &=& 1+\frac{32 \alpha  M}{\pi ^2 \alpha ^2+64 r^2}-\frac{4 M}{\pi  r} \tan ^{-1}\left(\frac{8 r}{\pi  \alpha }\right) +\frac{4 Q^2 }{r^2}\left(\frac{1}{\pi }\tan ^{-1}\left(\frac{8 r}{\pi  \alpha }\right)-\frac{8 \alpha  r}{\pi ^2 \alpha ^2+64 r^2}\right)^2+\frac{r^2}{\ell^2} 
\end{eqnarray}}
where $\alpha = 8\sqrt{\tfrac{\Theta}{\pi}}$. The profiles in Fig.~\ref{fig:mass-charge-profiles} illustrate the effect of NC Lorentzian smearing on the enclosed mass and charge. Both $m(r)/M$ and $q(r)/Q$ vanish at the origin, indicating that the effective sources are regular rather than point-like. As the radius increases, the enclosed values grow monotonically and approach unity asymptotically to the ADM mass $M$ and charge $Q$. The curves are therefore bounded by one, since the smeared distributions cannot contribute more than the total values defined at spatial infinity. Physically, the plot demonstrates that noncommutativity distributes the black hole’s mass and charge smoothly throughout the spacetime, interpolating between a regular core at $r=0$ and the standard RN-AdS behaviour at large distances. 

\section{Revisiting Thermodynamics and Criticality Perturbatively}\label{Sec:Thermodynamics and criticality with noncommutative parameter}

The modification in the metric as in  Eq.~\eqref{Working_metric_function1} preserves the overall RN structure while introducing corrections to the mass and charge functions that smoothly interpolate between the standard point-source limit and a regular core at small scales. Importantly, in the regime $\alpha \ll r_+$ (with $r_+$ the horizon radius), the smearing effects become exponentially suppressed, and the geometry asymptotically reduces to the usual RN black hole. Therefore, it is natural to interpret noncommutativity as a deformation of the RN spacetime. The metric~\eqref{Working_metric_function1} can be written as 
\begin{eqnarray}\label{Working_metric_function}
    f(r) = f_{RN}(r) + \alpha\frac{\left(M r^2-Q^2 r\right)}{r^4} +\alpha^2 \frac{Q^2}{4r^4}+ \mathcal{O}(\alpha^3) \ ,
\end{eqnarray}
where $f_{RN}(r)$ is the RN black hole metric. We can see the horizon radius numerically by equating Eq.~\eqref{Working_metric_function1} to zero. From Table~\ref{tab:Horizon_radius}, it is clear that non commutativity reduces the horizon radius.

\begin{table}[h!]
\centering
\begin{tabular}{|c|c|c|}
\hline
$\alpha$ & $r_+$ ($M=2,\; Q=1,\; \ell=1$) & $r_+$ ($M=1,\; Q=0.5,\; \ell=1$) \\ \hline
RN   & 1.2494 & 0.9290 \\ \hline
0.05 & 1.2397 & 0.9167 \\ \hline
0.10 & 1.2296 & 0.9037 \\ \hline
0.15 & 1.2194 & 0.8901 \\ \hline
0.20 & 1.2087 & 0.8756 \\ \hline
\end{tabular}
\caption{Event horizon radius $r_+$ for different values of $\alpha$ under two parameter sets.}
\label{tab:Horizon_radius}
\end{table}

From a thermodynamic perspective, we can treat the noncommutativity parameter as a perturbative parameter and study the thermodynamical quantities such that mass, temperature, equation of state, critical quantities, etc. can be consistently expanded perturbatively in powers of $\alpha$, with the zeroth-order terms reproducing the RN-AdS results and the higher-order terms capturing the deviations due to spacetime noncommutativity. This perturbative approach not only simplifies the algebraic structure of the criticality conditions but also provides a controlled way to quantify how the NC scale modifies these quantities. Now using the metric function as in Eq.~\eqref{Working_metric_function} and with the horizon condition \(f(r_+)=0\), the mass and the Hawking temperature are
\begin{eqnarray}\label{Mass_temp}
    M &=& M_{\rm RN}+\frac{\alpha  \left(\ell^2 r_+^2+r_+^4-\ell^2 Q^2\right)}{4 \ell^2 r_+^2}+\frac{\alpha ^2 \left(r_+^2+\ell^2\right)}{8 \ell^2 r_+} +\mathcal{O}(\alpha^3) \ , \\ \nonumber  T &=& T_{\rm RN}-\frac{\alpha  \left(r_+^4+\ell^2 r_+^2-3 \ell^2 Q^2\right)}{8 \pi  \ell^2 r_+^4}-\frac{\alpha ^2 \left(\ell^2 r_+^2+r_+^4+2 \ell^2 Q^2\right)}{16 \pi  \ell^2 r_+^5} +\mathcal{O}(\alpha^3) \ ,
\end{eqnarray}
where 
\begin{equation*}
    M_{\rm RN} = \frac{\ell^2 Q^2+\ell^2 r_+^2+r_+^4}{2 \ell^2 r_+} \qquad \text{and}\qquad  T_{\rm RN} = \frac{\ell^2 r_+^2 -\ell^2 Q^2 +3 r_+^4}{4 \pi  \ell^2 r_+^3} \ ,
\end{equation*}
are the RN black hole mass and Hawking temperature; also, we have suppressed the higher-order terms of $\alpha$. The Bekenstein–Hawking entropy reads
\begin{equation}\label{eq:S_bh}
    S_{BH} = \pi  r_+^2+\pi \, \alpha \, r_+ +\frac{1}{2} \alpha ^2 \log \left(\frac{r_+}{\ell}\right)+\mathcal{O}(\alpha^3) \ .
\end{equation}
The cosmological constant is identified as the pressure in the extended phase space~\cite{7}, which helps in writing the thermodynamic equation of state. The equation of state in terms of the specific volume is
\begin{eqnarray}\label{eq:P_of_T_rh}
    P(T,v) = P_{RN}(T,v)+\alpha  \left(\frac{T}{3 v^2}-\frac{16 Q^2}{3 \pi  v^5}+\frac{1}{3 \pi  v^3}\right)+\alpha ^2 \left(\frac{26 Q^2}{9 \pi  v^6}+\frac{4 T}{9 v^3}+\frac{4}{9 \pi  v^4}\right) \ ,
\end{eqnarray}
where $P_{RN}(T,v)$ is the equation of state for the RN case and already derived in \cite{39}. 


\begin{figure}[h!]
    \centering
    \includegraphics[width=0.45\textwidth, height=5cm]{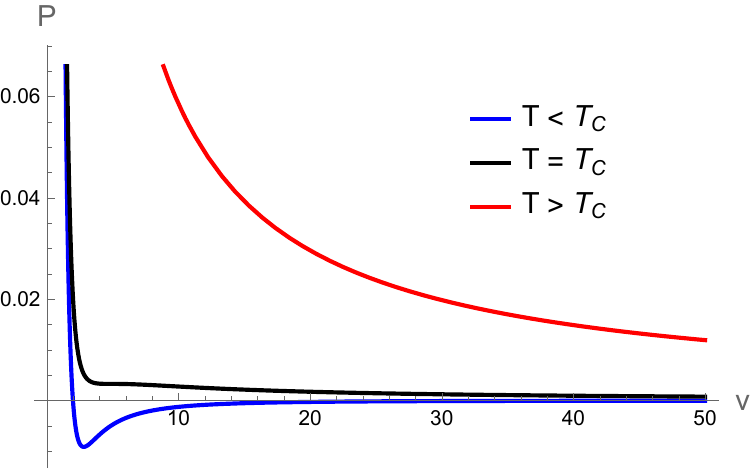}
    \includegraphics[width=0.45\textwidth, height=5cm]{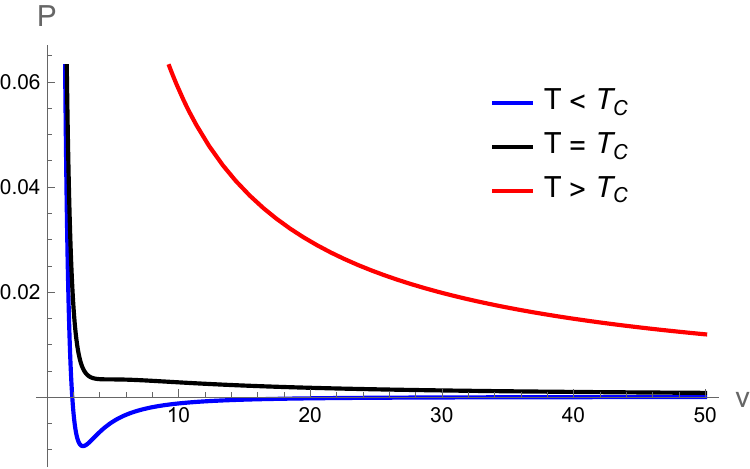}
    \vspace{5pt}
    \caption{{\bf Left:} Equation of state, i.e., $P-v$ diagram for fixed value of $\alpha=0.05$ and charge $Q=1$. {\bf Right:} $P-v$ diagram for fixed charge $Q=1$ for $\alpha=0.1$.}
    \label{Fig:P-vs-v}
\end{figure}

The critical point of the small/large black-hole phase transition is located by the inflection point conditions of the \(P\!-\!v\) isotherm Fig.~\ref{Fig:P-vs-v}, i.e., $\partial_v P=0$ and $\partial_v^2 P=0$. However, due to computational constraints, one cannot obtain closed-form expressions for critical quantities. The critical points of a black hole depend on a non-commutative parameter $\alpha$. At the critical temperature $T=T_c$, the oscillation collapses into a single inflection point, while for $T > T_c$ the isotherms are smooth and monotonic, corresponding to the absence of phase transitions.

To obtain the expressions for critical parameters valid perturbatively in \(\alpha\), i.e., we make the series ansatz up to \(\alpha^2\) as\footnote{We have assumed till \(\alpha^2\) because, during the derivation, we found that the pressure is corrected at \(\alpha^2\).}
\begin{equation}\label{v_CT_C Ansatz}
  v_c = v_0 + \alpha v_1 + \alpha^2 v_2 \;\;\;\;\;;\;\;\;\;\; T_c = T_0 + \alpha T_1 + \alpha^2 T_2 \ .  
\end{equation}
Now, we use the inflection point condition along with Eq.\eqref{v_CT_C Ansatz}, and we can solve this order by order. We start with the zeroth-order zero; the result is already matched with the result in~\cite{39} as $v_0 = 2\sqrt{6}\,Q$ and $T_0 = \tfrac{\sqrt{6}}{18\pi\,Q}$. Now, for the first order and by substituting $v_0, T_0$, solving the algebraic equation simultaneously, we have 
\begin{equation*}
    T_1 = -\frac{1}{72\pi\,Q^{2}}\qquad \text{and}\qquad v_1 = -1 \ .
\end{equation*}
Now, the second order with the substitution of $v_0, T_0, v_1, T_1$, and solving the algebraic equation simultaneously we have
\begin{equation*}
    T_2 = -\frac{13}{288 \sqrt{6} \pi  Q^3} \qquad \text{and}\qquad v_2 = \frac{13}{24 \sqrt{6} Q}  \ .
\end{equation*}
collecting them together, the critical parameters are 
\begin{eqnarray}\label{crit_par}
    v_c = v_c^{RN}-\alpha +\frac{13\sqrt{6} \alpha ^2}{144  Q} \,;\, T_c = T_c^{RN}-\frac{\alpha }{72 \pi  Q^2}-\frac{13\sqrt{6} \alpha ^2}{1728  \pi  Q^3} \,;\, P_c = P_c^{RN}-\frac{17 \alpha ^2}{6912 \pi  Q^4 } \ ,
\end{eqnarray}
where $v_c^{RN}$, $T_c^{RN}$, and $P_c^{RN}$ are critical quantities of the RN black hole~\cite{39}. Finally, at the critical point, we can compute the ratio $ \tfrac{P_c v_c}{T_c}$ as
\begin{eqnarray}
    \frac{P_c v_c}{T_c} = \frac{3}{8}-\frac{\sqrt{6} \alpha }{64 Q}-\frac{19 \alpha ^2}{768 Q^2}  \ .
\end{eqnarray}
The inclusion of the deformation parameter $\alpha$ modifies this universal value, thereby encoding the nontrivial imprint of the underlying spacetime structure on the critical behavior.

Before going for the main computation, we will briefly verify our perturbative computation by analyzing the Joule-Thomson (JT) expansion \cite{Okcu:2016tgt, Chabab:2018zix, RizwanCL:2018cyb, Cisterna:2018jqg, Mo:2018qkt, Lan:2018nnp, Mo:2018rgq} for a NC charged black hole. The JT process is characterized by curves of constant mass in the $T$-$P$ plane. Although these trajectories are technically constant-enthalpy (i.e., constant-mass) paths. The Joule-Thomson coefficient is defined as
\begin{equation}
\mu_J = \left( \frac{\partial T}{\partial P} \right)_M
      = \frac{1}{C_P} \left[\, T \left( \frac{\partial V}{\partial T} \right)_P - V \right] \ ,
\label{eq:JT_coefficient}
\end{equation}
from which the inversion temperature is obtained as
\begin{equation}
T_i = V \left( \frac{\partial T}{\partial V} \right)_P \ .
\label{eq:Ti_definition}
\end{equation}

The inversion temperature, expressed in terms of the horizon radius, is 
\begin{equation}
T_i =\frac{8 \pi  P r_+^4+3 Q^2-r_+^2}{12 \pi  r_+^3}+\frac{\alpha  \left(8 \pi  P r_+^4-33 Q^2+5 r_+^2\right)}{72 \pi  r_+^4}+\mathcal{O}(\alpha^2) \ .
\label{eq:Ti_rplus}
\end{equation}
Equating with the Hawking temperature as in Eq.~\eqref{Mass_temp} leads to the algebraic relation
\begin{equation}
\frac{-8 \pi  P r_+^4+3 Q^2-2 r_+^2}{6 \pi  r_+^3}+\frac{\alpha  \left(16 \pi  P r_+^4-30 Q^2+7 r_+^2\right)}{36 \pi  r_+^4}+\mathcal{O}(\alpha^2) = 0 \ .
\label{eq:relation_T_equals_Ti}
\end{equation}
Solving Eq.~\eqref{eq:relation_T_equals_Ti} for $r_+$ and selecting the physically meaningful root gives
\begin{equation}
r_+ = \frac{1}{2 \sqrt{2 \pi }}\sqrt{\frac{\mathcal{P_Q}-1}{P}}+\alpha  \left(-\frac{5}{24 \mathcal{P_Q}}-\frac{1}{3}\right)+\mathcal{O}(\alpha^2) \ .
\label{eq:rplus_solution}
\end{equation}
where $\sqrt{24 \pi P Q^2+1}= \mathcal{P_Q}$. Substituting this expression back into Eq.~\eqref{eq:Ti_rplus} yields the inversion temperature
\begin{equation}
\begin{aligned}
T_i &= \frac{\sqrt{P} \left(16 \pi  P Q^2-\mathcal{P_Q}+1\right)}{\sqrt{2 \pi } \left(\mathcal{P_Q}-1\right)^{3/2}}-\frac{\alpha  \left(192 \pi \mathcal{P_Q} P Q^2+6 \mathcal{P_Q}^2-5+\mathcal{P_Q}\right)}{288 \pi\mathcal{P_Q}  Q^2} \ .
\end{aligned}
\label{eq:Ti_final}
\end{equation}
Equation \eqref{eq:Ti_final} shows that the inversion temperature depends on the non-commutative parameter $\alpha$, and the electric charge $Q$. Finally, setting inversion pressure to zero yields the minimal inversion temperature
\begin{equation}
T_i^{\min} = \frac{1}{6 \sqrt{6} \pi  Q}-\frac{\alpha }{144 \pi  Q^2} +\mathcal{O}(\alpha^2) \ .
\label{eq:Timin}
\end{equation}
The ratio of the minimal inversion temperature to the critical temperature, as in Eq.~\eqref{crit_par}, then evaluates to
\begin{equation}
\frac{T_i^{\min}}{T_c} = \frac{1}{2} \ .
\label{eq:ratio}
\end{equation}
Remarkably, this ratio remains unchanged by the presence of the NC parameter $\alpha$. The Joule-Thomson expansion of the NC black hole therefore shares the same universality class as the RN-AdS black hole, although $\alpha$ modifies the absolute values of the inversion and critical temperatures.

\section{Topological Charge in Bulk}\label{Sec:Topological Study}

The topological characterization of black hole thermodynamics is formulated through an off-shell extension of the Helmholtz free energy. For a black hole of mass $M$ and entropy $S$ immersed in a thermal reservoir of fixed inverse temperature $\tau^{-1}$, the generalized free energy is defined as
\begin{equation}\label{GFE}
    \mathcal{F} = M - \frac{S}{\tau} \ ,
\end{equation}
which reduces to the usual thermodynamic potential when $\tau^{-1}=T_H$. This off-shell construction provides a natural link between the Euclidean path-integral description and a topological classification of black hole phases.

To extract the associated topological information, an auxiliary angular parameter $\Theta\in(0,\pi)$ is introduced and a two-component vector field is defined as
\begin{equation}\label{phi}
    \vec{\phi} = \left( \frac{\partial \mathcal{F}}{\partial r_+},\; -\cot\Theta\,\csc\Theta \right) .
\end{equation}
The physically admissible black hole configurations correspond to the zeros of the radial component of $\vec{\phi}$. Each zero behaves as a topological defect in the $(r_+,\Theta)$ plane and is characterized by an integer winding number. The sum of the winding numbers over all defects defines the global topological charge, which serves as a robust invariant encoding the overall thermodynamic phase structure of the black hole.

To each configuration of $\vec{\phi}$ one may associate an identically conserved current,
\begin{equation}\label{j}
    j^{\mu} = \frac{1}{2\pi}\,\epsilon^{\mu\nu\rho}\, \epsilon_{ab}\, \partial_{\nu} n^{a}\, \partial_{\rho} n^{b} \ ,
\end{equation}
where the unit vector $n^a$ is defined through
\begin{equation}\label{n}
    n^1=\frac{\phi^1}{\phi},\qquad n^2=\frac{\phi^2}{\phi},\qquad \phi = |\vec{\phi}| \ .
\end{equation}
Because $j^\mu$ is constructed purely from antisymmetric combinations, it vanishes everywhere except at points where $\vec{\phi}$ becomes zero, thereby localizing nontrivial topology entirely at the defects. One can easily compute the temporal component of the conserved current as,
\begin{equation}
    j^0 = \frac{1}{\pi} \left(\partial_1 n^1\,\partial_2 n^2 - \partial_2 n^1\,\partial_1 n^2 \right) \ .
\end{equation}
Within a region $D$ where $\phi \neq 0$, both components of $\vec{\phi}$ are smooth and one may rewrite $j^0$ in the divergence form
\begin{equation}
    j^0 = \frac{1}{\pi}\left( \partial_1 \mathcal{Q} - \partial_2 \mathcal{P} \right), \qquad \mathcal{Q} = n^1 \partial_2 n^2,\quad \mathcal{P} = n^1 \partial_1 n^2 \ .
\end{equation}
Applying Green’s theorem, we have
\begin{equation}
    0 = \int_D j^0\, d^2 x = \frac{1}{\pi} \oint_{\partial D} \left( \mathcal{P}\, dx^1 + \mathcal{Q}\, dx^2 \right) = \frac{1}{\pi}  \oint_{\partial D} n^1\, dn^2 \ .
\end{equation}
Let $C$ be a closed contour enclosing all zeros of $\vec{\phi}$. The quantity
\begin{equation}\label{W}
    W = \frac{1}{\pi} \oint_C n^1\, dn^2 = \frac{1}{\pi} \sum_{i=1}^{N} \oint_{c_i} n^1\, dn^2
\end{equation}
defines a topological invariant, where each $c_i$ is a small loop surrounding a single zero. To evaluate each contribution, consider $\vec{\phi}=(f(x),g(y))$ with $f(x_0)=g(y_0)=0$ and choose $g'(y_0)=1$ for convenience. For a circular contour of radius $\varepsilon$ around $(x_0,y_0)$, one finds the leading behavior
\begin{equation*}
    f(x(t)) = \varepsilon f'(x_0)\cos t + \mathcal{O}(\varepsilon) \qquad ; \qquad g(y(t)) = \varepsilon\sin t + \mathcal{O}(\varepsilon) \ .
\end{equation*}
Taking $\varepsilon \to 0$ yields the local index
\begin{equation}
    \oint_{c_i} n^1\, dn^2 = \pi\, \mathrm{sgn}\!\left(f'(x_0)\right)
\end{equation}
whenever $f'(x_0)\neq 0$. Summing over all such points gives
\begin{equation}\label{W1}
    W  =\sum_{i=1}^{N} w_i = \sum_{i=1}^{N} \mathrm{sgn}\!\left. \left( \frac{\partial^2 \mathcal{F}}{\partial r_+^2} \right) \right|_{r_+=r_i} \ ,
\end{equation}
where the $r_i$ solve $\partial_{r_+}\mathcal{F}=0$ and $w_i$ is the winding number. The sign function $\mathrm{sgn}(x)$ takes the value $+1, -1$ and $0$ for $x>0$, $x<0$ and $x=0$ respectively. It is worth emphasizing that the sign of the winding number carries direct physical information. In particular, it has been shown in~\cite{Anand:2025cer} that the sign of the winding number $w_i$ distinguishes thermodynamically stable and unstable branches of black-hole solutions. Specifically, a positive winding number ($w_i=+1$) corresponds to a locally stable equilibrium branch, whereas a negative winding number ($w_i=-1$) signals thermodynamic instability. Our result in Eq.~\eqref{W1} is fully consistent with this interpretation, as the sign of $\partial_{r_+}^2\mathcal{F}$ simultaneously determines both the local curvature of the generalized free energy and the associated thermodynamic stability.

For the charged NC black hole, the generalized free energy is 
\begin{eqnarray}
    \mathcal{F} &=& \frac{8 \pi  P r_+^4 \tau +Q^2 \tau -2 \pi  r_+^3+r_+^2 \tau }{2 r_+ \tau }+\alpha  \left(\frac{8 \pi  P r_+^4-Q^2+r_+^2}{4 r_+^2}-\frac{\pi  r_+}{\tau }\right) \nonumber \\
    && \qquad\qquad + \frac{\alpha ^2 \left(8 \pi  P r_+^2 \tau -4 \pi  r_+ \log (r_+)+\tau \right)}{8 r_+ \tau } +\mathcal{O}(\alpha^3) \ .
\end{eqnarray}
Now, the two-component field read
\begin{eqnarray}\label{phi}
    \phi^1 &=&\frac{\left(96 \pi  P r_+^2-\frac{4 Q^2}{r_+^2}-\frac{16 \pi  r_+}{\tau }+4\right)}{8} + \frac{ \alpha \left(32 \pi  P r_++\frac{4 Q^2}{r_+^3}-\frac{8 \pi }{\tau }\right)}{8} +\frac{\alpha^2\,\left(8 \pi  P+\frac{-4 \pi  r_+-\tau }{r_+^2 \tau }\right)}{8}  \nonumber \\
    \phi^2 &=& -\cot\theta\,\csc\theta\,.
\end{eqnarray}
By solving the constraint $\partial_{r_+}\mathcal{F}(r_+) = \phi^1=0$ reveals 
\begin{eqnarray}\label{Tau_nc}
    \tau &=& \frac{4 \pi  r_+^3}{24 \pi  P r_+^4-Q^2+r_+^2}+\frac{2 \pi  \alpha  r_+^2 \left(8 \pi  P r_+^4-3 Q^2+r_+^2\right)}{\left(24 \pi  P r_+^4-Q^2+r_+^2\right)^2} \nonumber \\
    &&+\frac{\pi  \alpha ^2 r_+ \left(256 \pi ^2 P^2 r_+^8-8 \pi  P Q^2 r_+^4+48 \pi  P r_+^6+7 Q^4-5 Q^2 r_+^2+2 r_+^4\right)}{\left(24 \pi  P r_+^4-Q^2+r_+^2\right)^3} \ .
\end{eqnarray}
To evaluate the winding number associated with the zero of $\partial \mathcal{F}/\partial r_+$, we make use of
Eq.~\eqref{W1}. For the charge NC black string, the second derivative of the free energy takes the form
\begin{equation}\label{Fpp}
\left.\frac{\partial^2 \mathcal{F}}{\partial r_+^2}\right|_{r_+=r_i} = \frac{\alpha  \left(8 \pi  P r_+^4-3 Q^2\right)}{2 r_+^4}+\frac{24 \pi  P r_+^4 \tau +Q^2 \tau -2 \pi  r_+^3}{r_+^3 \tau }+\frac{\alpha ^2 (2 \pi  r_++\tau )}{4 r_+^3 \tau } \ .
\end{equation}

From Eq.~\eqref{Tau_nc}, it is clear that it is not analytically tractable to compute the critical radius \(r_i\) for fixed $\tau$. We should take another path, i.e., the numerical one. We will compute for different values of $\alpha$ up to second order. The discussion is organized in order of the deformation parameter $\alpha$.

\subsection*{For $\alpha = 0.05$}

\subsubsection*{Up to first order}
For up to the first order, we have 
\begin{eqnarray}\label{F_at_alpha}
    \mathcal{F} = \frac{8 \pi  P r_+^4 \tau +Q^2 \tau -2 \pi  r_+^3+r_+^2 \tau }{2 r_+ \tau }+\alpha  \left(\frac{8 \pi  P r_+^4-Q^2+r_+^2}{4 r_+^2}-\frac{\pi  r_+}{\tau }\right) \ .
\end{eqnarray}
By solving $\partial_{r_+}\mathcal{F}|_{r_+=r_i}=0$, we can compute the $\tau$ as 
\begin{eqnarray}\label{Tau_at_alpha}
    \tau =  \frac{2 \pi  r_i^3 (\alpha +2 r_i)}{8 \pi  P r_i^4 (\alpha +3 r_i)+Q^2 (\alpha -r_i)+r_i^3} \ .
\end{eqnarray}
It is not possible to determine $r_i$ analytically for a fixed value of $\tau$. By fixing $\tau=40$, $P<P_c$, and using numerical techniques, we can compute the $r_i$. Finally, to compute the winding number, we have to check the sign of $\partial^2_{r_+}\mathcal{F}|_{r_+=r_i}$. The second order derivative of generalized free energy as in Eq.~\eqref{F_at_alpha} is
\begin{eqnarray}\label{d2F_at_alpha}
   \frac{\partial^2 \mathcal{F}}{\partial^2 r_+}\Bigg|_{r_+=r_i} = 4 \pi  P (\alpha +6 r_i)+\frac{Q^2 (2 r_i-3 \alpha )}{2 r_i^4}-\frac{2 \pi }{\tau }
\end{eqnarray}

From the table~\ref{tab:combined_stability}, the global topological charge can be computed using Eq.~\eqref{W1} as 
\begin{eqnarray}
    W = 0 \ .
\end{eqnarray}
Now, for $\tau =20 $, we have two real roots of Eq.~\eqref{Tau_at_alpha}, computing the second derivative~\eqref{d2F_at_alpha} at these $r_i's$ value. From table~\ref{tab:combined_stability}, again the global topological charge is zero.

For $P>P_c$, for any values of, $\tau$ we have solved the Eq.~\eqref{Tau_at_alpha}. By fixing $\tau=40$, we have two real values for $r_i$. With them we have computed the second derivative sign at those $r_i's$ and that results in the winding number at those $r_i's$. Table~\ref{tab:combined_stability}, contains the detailed information for $P>P_c$.
\begin{table}[h!]
\centering
\begin{tabular}{|c|c|c|c|c|}
\hline
Case & $r_i$ & $\partial^2_{r_+}\mathcal{F}|_{r_+=r_i}$ & $w_i$ & Stability \\ \hline

$P<P_c,\; Q=1,\; \tau=40$ & 0.0501 & $-$ve & $-1$ & Unstable \\ \hline
$P<P_c,\; Q=1,\; \tau=40$ & 1.2317  & $+$ve & $+1$ & Stable   \\ \hline
$P<P_c,\; Q=1,\; \tau=40$ & 3.7094  & $-$ve & $-1$ & Unstable \\ \hline
$P<P_c,\; Q=1,\; \tau=40$ & 13.6269  & $+$ve & $+1$ & Stable   \\ \hline
$P<P_c,\; Q=1,\; \tau=20$ & 0.0501  & $-$ve & $-1$ & Unstable \\ \hline
$P<P_c,\; Q=1,\; \tau=20$ & 33.8024  & $+$ve & $+1$ & Stable   \\ \hline
$P>P_c,\; Q=1,\; \tau=40$ & 0.0501 & $-$ve & $-1$ & Unstable \\ \hline
$P>P_c,\; Q=1,\; \tau=40$ & 0.4087  & $+$ve & $+1$ & Stable   \\ \hline

\end{tabular}
\caption{Thermodynamic stability and winding numbers for different pressure regimes and Euclidean time periods.}
\label{tab:combined_stability}
\end{table}
\subsubsection*{Up to second order}

Now, retaining terms up to second order in the deformation parameter $\alpha$, the generalized free energy takes the form
\begin{eqnarray}\label{F_at_alpha2}
\mathcal{F} &=& \frac{8 \pi  P r_+^4 \tau +Q^2 \tau -2 \pi  r_+^3+r_+^2 \tau }{2 r_+ \tau }+\alpha  \left(\frac{8 \pi  P r_+^4-Q^2+r_+^2}{4 r_+^2}-\frac{\pi  r_+}{\tau }\right)\nonumber \\
&&+\frac{\alpha ^2 \left(8 \pi  P r_+^2 \tau -4 \pi  r_+ \log (r_+)+\tau \right)}{8 r_+ \tau } \ .
\end{eqnarray}
The stationary points of the free energy are obtained from the condition $\partial_{r_+}\mathcal{F}=0$. Solving this relation for $\tau$ at $r_+=r_i$, yields
\begin{eqnarray}\label{Tau_at_alpha2}
\tau =  \frac{4 \pi  r_+^2 \left(\alpha ^2+4 r_+^2+2 \alpha  r_+\right)}{96 \pi  P r_+^5+32 \pi  \alpha  P r_+^4+r_+^3 \left(8 \pi  \alpha ^2 P+4\right)+4 Q^2 (\alpha -r_+)-\alpha ^2 r_+} \ .
\end{eqnarray}

Since Eq.~\eqref{tab:combined_alpha2} cannot be inverted analytically to obtain $r_i$ for a fixed value of $\tau$, we proceed numerically. For fixed $\tau$ and pressure below the critical value ($P<P_c$), multiple real solutions for $r_i$ are obtained. In particular, for $\tau=40$ and $Q=1$, four real extrema appear, as summarized in Table~\ref{tab:combined_alpha2}. For a smaller value $\tau=20$, only two real extrema are present, as shown in Table~\ref{tab:combined_alpha2}.

To determine the local stability of each extremum and the associated winding number, we evaluate the second derivative of the free energy. From Eq.~\eqref{F_at_alpha2}, we obtain
\begin{eqnarray}\label{d2F_at_alpha2}
\frac{\partial^2 \mathcal{F}}{\partial r_+^2}\Bigg|_{r_+=r_i}
=\frac{8 \pi  r_i^4 (2 P \tau  (\alpha +6 r_i)-1)+Q^2 \tau  (4 r_i-6 \alpha )+2 \pi  \alpha ^2 r_i^2+\alpha ^2 r_i \tau }{4 r_+^4 \tau } \ .
\end{eqnarray}
The sign of Eq.~\eqref{d2F_at_alpha2} determines the local stability of the corresponding branch and fixes the winding number $w_i=\pm1$. Summing over all extrema using Eq.~\eqref{W1}, we find that the total topological charge vanishes,
\begin{eqnarray}
W = 0 \ ,
\end{eqnarray}
for both $\tau=40$ and $\tau=20$ in the subcritical pressure regime. We now turn to the supercritical case $P>P_c$. In this regime, Eq.~\eqref{Tau_at_alpha} admits two real solutions for $r_i$ for any fixed value of $\tau$. For definiteness, we choose $\tau=40$. The corresponding extrema, their stability properties, and associated winding numbers are listed in Table~\ref{tab:combined_alpha2}. Once again, applying Eq.~\eqref{W1}, we find that the total topological charge in the supercritical regime also vanishes,
\begin{eqnarray}
W = 0 \ .
\end{eqnarray}

\begin{table}[h!]
\centering
\begin{tabular}{|c|c|c|c|c|}
\hline
Case & $r_i$ & $\partial^2_{r_+}\mathcal{F}|_{r_+=r_i}$ & $w_i$ & Stability \\ \hline

$P<P_c,\; Q=1,\; \tau=40$ & 0.0501   & $-$ve & $-1$ & Unstable \\ \hline
$P<P_c,\; Q=1,\; \tau=40$ & 1.2324   & $+$ve & $+1$ & Stable   \\ \hline
$P<P_c,\; Q=1,\; \tau=40$ & 3.7088   & $-$ve & $-1$ & Unstable \\ \hline
$P<P_c,\; Q=1,\; \tau=40$ & 13.6270  & $+$ve & $+1$ & Stable   \\ \hline
$P<P_c,\; Q=1,\; \tau=20$ & 0.0500874 & $-$ve & $-1$ & Unstable \\ \hline
$P<P_c,\; Q=1,\; \tau=20$ & 33.8024   & $+$ve & $+1$ & Stable   \\ \hline
$P>P_c,\; Q=1,\; \tau=40$ & 0.0501023 & $-$ve & $-1$ & Unstable \\ \hline
$P>P_c,\; Q=1,\; \tau=40$ & 0.408665  & $+$ve & $+1$ & Stable   \\ \hline
\end{tabular}
\caption{Thermodynamic stability analysis for different pressure regimes and Euclidean periods.}
\label{tab:combined_alpha2}
\end{table}

\subsection*{Analysis at $\alpha = 0.1$}

\subsubsection*{Linear order in $\alpha$}

Retaining only terms up to $\mathcal{O}(\alpha)$, the generalized free energy is expressed as in Eq.~\eqref{F_at_alpha}. Using the equilibrium configurations corresponds to extrema of $\mathcal{F}$ with respect to the horizon radius. Imposing the stationary condition $\partial_{r_+}\mathcal{F}=0$ and evaluating it at $r_+=r_i$ allows one to solve for the auxiliary parameter $\tau$ as in Eq.~\eqref{Tau_at_alpha}.

For a chosen value of $\tau$, Eq.~\eqref{Tau_at_alpha} cannot be inverted analytically to determine $r_i$. Consequently, the locations of the extrema are obtained numerically again. In the subcritical pressure regime ($P<P_c$), multiple real solutions are found. Specifically, for $Q=1$ and $\tau=40$, four distinct extrema arise, while for $\tau=20$ only two real solutions persist. The corresponding stability properties are summarized below in Table~\ref{tab:combined_alpha01}.

The local stability of each extremum is determined by the second derivative of the free energy as in Eq.~\eqref{d2F_at_alpha}. The sign of Eq.~\eqref{d2F_at_alpha} fixes the winding number $w_i=\pm1$. Summing over all extrema according to Eq.~\eqref{W1}, we find that the total topological charge vanishes,
\begin{eqnarray}
W = 0 \ ,
\end{eqnarray}
for both $\tau=40$ and $\tau=20$ in the subcritical regime. For pressures exceeding the critical value ($P>P_c$), the structure can be simplified. In this case, Eq.~\eqref{Tau_at_alpha} admits only two real extrema for any fixed $\tau$. Choosing $\tau=40$, the resulting configurations and their stability
properties are listed in Table~\ref{tab:combined_alpha01}. The corresponding total topological charge again evaluates to
\begin{eqnarray}
W = 0 \ .
\end{eqnarray}

\begin{table}[h!]
\centering
\begin{tabular}{|c|c|c|c|c|}
\hline
Case & $r_i$ & $\partial^2_{r_+}\mathcal{F}$ & $w_i$ & Stability \\ \hline
$P<P_c,\; Q=1,\; \tau=40$ & 0.10098  & $-$ve & $-1$ & Unstable \\ \hline
$P<P_c,\; Q=1,\; \tau=40$ & 1.20414  & $-$ve & $-1$ & Unstable \\ \hline
$P<P_c,\; Q=1,\; \tau=40$ & 3.66947  & $+$ve & $+1$ & Stable   \\ \hline
$P<P_c,\; Q=1,\; \tau=40$ & 13.6493  & $+$ve & $+1$ & Stable   \\ \hline
$P<P_c,\; Q=1,\; \tau=30$ & 0.100964 & $-$ve & $-1$ & Unstable \\ \hline
$P<P_c,\; Q=1,\; \tau=30$ & 20.9796  & $+$ve & $+1$ & Stable   \\ \hline
$P>P_c,\; Q=1,\; \tau=40$ & 0.101017 & $+$ve & $+1$ & Stable   \\ \hline
$P>P_c,\; Q=1,\; \tau=40$ & 0.661782 & $-$ve & $-1$ & Unstable \\ \hline
\end{tabular}
\caption{Thermodynamic stability and winding numbers for different pressure regimes and Euclidean periods.}
\label{tab:combined_alpha01}
\end{table}

\subsubsection*{Quadratic order in $\alpha$}

We now incorporate second-order corrections in the deformation parameter. Including terms up to $\mathcal{O}(\alpha^2)$, the generalized free energy takes the form as in Eq.~\eqref{F_at_alpha2}. The equilibrium condition $\partial_{r_+}\mathcal{F}=0$ now leads to Eq.~\eqref{Tau_at_alpha2}. As in the linear case, the extrema must be determined numerically. The resulting solutions for subcritical and supercritical pressures closely mirror the first-order structure, with only quantitative shifts in the extrema's locations. The stability analysis follows from the second derivative~\eqref{d2F_at_alpha2}. Evaluating Eq.~\eqref{d2F_at_alpha2} at each extremum fixes the winding numbers, and the total topological charge is again found to vanish,
\begin{eqnarray}
W = 0 \ ,
\end{eqnarray}
both below and above the critical pressure. It is to be noted that the number of roots of eq.~\eqref{Tau_nc} depends on $\alpha$, hence we expect that the number of black holes for a given $\tau$ and fixed charge should depend on the value of the NC parameter $\alpha$. Nevertheless, the total winding number $W$ is fixed and hence independent of $\alpha$ as it contains the topological information of the black hole class. By invoking non-commutativity, branches of a black hole emerge, which were absent in the RN black holes case~\cite{a19}, thus making the NC black hole of an altogether different class from RN black holes. 

\section{Topological Charge in Dual CFT}\label{Sec:Topological Charge in Dual CFT}

In this section, we compute the topological charge in the dual boundary case. For this, we start by writing the metric as in Eq.~\eqref{Working_metric_function} to see that the apparent asymmetry in the appearance of Newton's constant \(G\) between the classical RN–AdS term and the NC correction has a clear physical origin. The classical term descends directly from Einstein's field equations, where \(G\) is essential for coupling matter to geometry. In contrast, the NC corrections arise from a fundamentally different source: a deformation of spacetime itself, encoded in the NC algebra of coordinates. Now, using the metric function, including $G$ and imposing the horizon condition \( f(r_+) = 0 \), we derive the mass of the black hole is
    \begin{equation}
    M=\frac{r_{+}}{2G}\!\left(1+\frac{r_{+}^{2}}{\ell^{2}}+\frac{GQ^{2}}{r_{+}^{2}}\right) -\frac{\alpha}{4G}\!\left(1+\frac{r_{+}^{2}}{\ell^{2}}-\frac{GQ^{2}}{r_{+}^{2}}\right) +\frac{\alpha^{2}}{8G r_{+}}\!\left(1+\frac{r_{+}^{2}}{\ell^{2}}\right)+\mathcal{O}(\alpha^{3}) \ . \nonumber
    \end{equation}
Also, the Hawking temperature $T$ is 
    \begin{multline}
    T=\frac{1}{4\pi r_{+}}\!\left(1+\frac{3r_{+}^{2}}{\ell^{2}}-\frac{GQ^{2}}{r_{+}^{2}}\right)-\frac{\alpha}{8\pi r_{+}^{2}}\!\left(1+\frac{r_{+}^{2}}{\ell^{2}}-\frac{3GQ^{2}}{r_{+}^{2}}\right)-\frac{\alpha^{2}}{16\pi r_{+}^{3}}\!\left(1+\frac{r_{+}^{2}}{\ell^{2}}+\frac{2GQ^{2}}{r_{+}^{2}}\right) \ . \nonumber 
    \end{multline}
Following the extended boundary thermodynamics~\cite{Ahmed:2023snm}, we map the bulk condition to a dual conformal field theory (CFT). The dual CFT resides on the conformal boundary of AdS$_4$. We introduce a dimensionless conformal factor $\omega$. The boundary metric is:
    \[
    ds^{2}= \omega^{2}\bigl(-dt^{2}+\ell^{2}d\Omega_{2}^{2}\bigr) \ ,
    \]
where $\omega = R/\ell$, and $R>0$ is the curvature radius of the boundary, treated as a free thermodynamic variable. The spatial volume of the boundary is $\mathcal{V}=4\pi R^{2}$. The holographic dictionary relating bulk gravitational quantities to boundary thermodynamic variables,
\begin{equation}
\tilde{E}=M\frac{\ell }{R}=\frac{M}{\omega}, \quad \tilde{S}=S,  \quad \tilde{T}=T\frac{\ell }{R}=\frac{T}{\omega}, \quad\tilde{Q}=Q\frac{\ell }{\sqrt{G}}=2Q\sqrt{C}, \quad \tilde{\phi}=\frac{\phi }{R},
\label{dictonary}
\end{equation}
where tilde quantities represent boundary CFT variables: \(\tilde{E}\) (energy), \(\tilde{T}\) (temperature), \(\tilde{S}\) (entropy), \(\tilde{\phi}\) (electrostatic potential), and \(\tilde{Q}\) (charge). For the boundary theory with variable central charge $c$ and volume $\mathcal{V}$, the first law reads
    \begin{equation}
    d\tilde{E}= \tilde{T}\,d\tilde{S}+ \tilde{\Phi}\,d\tilde{Q}+ \mathcal{A}d\alpha+\mu_{c}\,dc - P d\mathcal{V} \ .
    \end{equation}
Here, $P$ is the boundary pressure (conjugate to $\mathcal{V}$), and $\mu_{c}$ is the chemical potential (conjugate to $c$). The corresponding Euler relation, dual to the Smarr relation, is:
    \begin{equation}
    \tilde{E}= \tilde{T}\tilde{S}+ \tilde{\Phi}\tilde{Q}+\mathcal{A}\alpha +\mu_{c}c \ .
    \end{equation}
The $P \mathcal{V}$ term is absent and shows consistency with CFT. Using the dictionary~\eqref{dictonary} with $\omega$, we can write the boundary thermodynamic quantity in the terms of $r_{+}$, $\tilde{Q}$, and $c$ as
   \begin{align}
    \tilde{E}(r_+) &= \frac{2\sqrt{c}}{R} \Bigg[\frac{r_+}{2}\left(1 + \frac{\tilde{Q}^2}{r_+^2} + \frac{r_+^2}{4c}\right) - \frac{\alpha}{4}\left(1 - \frac{\tilde{Q}^2}{r_+^2} + \frac{r_+^2}{4c}\right) + \frac{\alpha^2}{8r_+}\left(1 + \frac{r_+^2}{4c}\right) \Bigg], \\[10pt]
    \tilde{T}(r_+) &= \frac{2\sqrt{c}}{R} \Bigg[ 
        \frac{1}{4\pi r_+}\left(1 + \frac{3r_+^2}{4c} - \frac{\tilde{Q}^2}{r_+^2}\right) - \frac{\alpha}{8\pi r_+^2}\left(1 + \frac{r_+^2}{4c} - \frac{3\tilde{Q}^2}{r_+^2}\right) \nonumber \\
        & \qquad \qquad - \frac{\alpha^2}{16\pi r_+^3}\left(1 + \frac{r_+^2}{4c} + \frac{2\tilde{Q}^2}{r_+^2}\right) \Bigg],  \qquad  \tilde{S}(r_+) = S(r_+) \ .
\end{align}
The boundary pressure $P= \tilde{E}/(2\mathcal{V})$ follows from Euler's equation~\cite{Ahmed:2023snm}. We now apply the topological method to the boundary CFT described above. We work in the fixed $(\tilde{Q},\mathcal{V},c)$ ensemble, i.e., the canonical ensemble for the boundary theory. Following Eq.~\eqref{GFE}, we introduce an off-shell generalized free energy
    \begin{equation}
    \mathcal{\tilde{F}} = \tilde{E} - \frac{\tilde{S}}{\tilde{\tau}} \ ,
    \end{equation}
where $\tilde{\tau}>0$ is an auxiliary parameter with dimensions of inverse temperature at the boundary. The physical, on-shell condition is $\tilde{\tau}= 1/\tilde{T}$.  We compute the components of vector field $\vec{\phi}$ on the $(r_{+},\Theta)$ plane, with $\Theta\in(0,\pi)$ as and from the zero point constraint $\phi^{1}=0$ reveals the form of $\tilde{\tau}$ as 
\begin{equation}
    \tilde{\tau}(r_+) = \frac{\pi r_+ R}{\sqrt{c}} \left[ \frac{1}{2} - \frac{\tilde{Q}^2}{2r_+^2} + \frac{3r_+^2}{8c} - \frac{\alpha \tilde{Q}^2}{2r_+^3} - \frac{\alpha r_+}{8c} - \frac{\alpha^2}{8r_+^2} + \frac{\alpha^2}{32c} \right]^{-1}.
\end{equation}
Since $\phi^{2}=0$ only at $\Theta=\pi/2$, all zero points lie on this line. Therefore, the zero points of $\vec{\phi}$ correspond exactly to the black hole solutions (phases of the CFT) at a given temperature $\tilde{T}=1/\tilde{\tau}$. As each zero point $(r_{h}=r_{i},\Theta=\pi/2)$ is assigned a winding number $w_{i}$:
    \begin{equation}
    w_{i}= \operatorname{sgn}\!\left(\frac{\partial\phi^{1}}{\partial r_{+}}\right)_{r_{+}=r_{i}} = \operatorname{sgn}\!\left(\frac{\partial^{2}\mathcal{\tilde{F}}}{\partial r_{+}^{2}}\right)_{r_{+}=r_{i}} \ .
    \end{equation}
 The total topological charge $W$ for a given set of parameters $(\tilde{\tau},\tilde{Q},\mathcal{V},c)$ is the sum of winding numbers of all enclosed zero points:

For the Reissner-Nordström case, the topological charge is $W=+1$~\cite{Zhang:2023uay}. Now, once the NC correction terms in $\tilde{T}(r_+)$ become dominant for very small $r_+$. Specifically, the term
    \[
    -\frac{\alpha}{8\pi R\sqrt{c}}\!\left(-\frac{3\tilde{Q}^{2}}{r_+^{4}}\right)
    =+\frac{3\alpha \tilde{Q}^{2}}{8\pi R\sqrt{c}\,r_+^{4}}
    \]
is positive and diverges as $r_+\to0$. This positive divergence causes $\tilde{T}(r_+)$ to become positive again at an extremely small value of $r_+$, creating a new branch of black hole solutions. This introduces a fourth real positive root, $r_+^{(0)}$, which is smaller than all roots of the commutative case ($r_+^{(0)}\ll r_+^{(1)}$). On this new branch, the temperature decreases with increasing $r_+$ (i.e., $\partial\tilde{T}/\partial r_+<0$) due to the dominant $1/r_+^{4}$ behavior. Therefore, this new zero point has a winding number:
    \begin{equation}
    w^{(0)}=-1 \ .
    \end{equation}
For $\alpha>0$, and for a range of $\tilde{\tau}$ where all branches coexist, the equation $\tilde{\tau}=1/\tilde{T}(r_+)$ now yields four real positive roots:
    \[
    r_+^{(0)}<r_+^{(1)}<r_+^{(2)}<r_+^{(3)},
    \]
    with winding numbers $w^{(0)}=-1$, $w^{(1)}=+1$, $w^{(2)}=-1$, $w^{(3)}=+1$. The total topological charge is
    \begin{equation}
    W_{\alpha>0}=0 \ .
    \end{equation}
This result shows:
\begin{itemize}
    \item It is independent of the specific positive values of $\tilde{Q}$, $R$, $c$, and $\alpha$, provided $\alpha>0$.
    \item It holds for both subcritical ($P<P_c$) and supercritical ($P>P_c$) boundary pressures. In the supercritical case, the intermediate roots $r_+^{(1)}$ and $r_+^{(2)}$ may merge and disappear, but the very small root $r_+^{(0)}$ and the large stable root $r_+^{(3)}$ persist, typically with $w^{(0)}=-1$ and $w^{(3)}=+1$, still summing to $W=0$.
    \item It is consistent to all perturbative orders in $\alpha$; the existence of the new root is guaranteed by the leading-order NC correction.
\end{itemize}

This represents a change in topological class. The non-zero value $W=+1$ characterizes holographic systems whose gravitational duals possess a classical, singular inner structure. The value $W=0$ characterizes systems whose gravitational duals incorporate a fundamental minimal length scale (here via noncommutativity) that regularizes the singularity and modifies the deep UV thermodynamics.  The boundary CFT thermodynamics, through this topological invariant, encodes profound information about the quantum gravitational nature of the bulk spacetime. The transition from $W=+1$ to $W=0$ upon turning on $\alpha$ is a direct topological signature of the singularity resolution in the bulk, manifesting in the boundary theory as the appearance of a novel, unstable branch of ``microscopic'' black hole states. 


\section{Binary Black Hole Merger and Bounds on the Final Mass}\label{Sec: Binary Black Hole Merger}
In this section, we consider the head-on collision of two charged black holes incorporating NC correction terms. The NC parameter \(\alpha\), defined in terms of the parameter \(\theta\) as \(\alpha = 8\sqrt{\theta/\pi}\), is analyzed in relation to the final black-hole mass after the merger. This allows us to obtain a lower bound on the final mass imposed by the second law of black-hole thermodynamics.

This analysis emphasizes the role of NC corrections in modifying thermodynamic quantities, particularly the entropy and the final black-hole mass. It also provides a qualitative understanding of how variations in the NC parameter affect the final mass and the energy radiated through gravitational waves. These results differ from the predictions of classical general relativity. Although the NC correction parameter is currently beyond direct experimental detection, this study may provide useful theoretical guidance for identifying possible observational signatures of such corrections in future investigations. As we employ the perturbative expressions for the mass and entropy up to second order in $\alpha$, given in Eqs.~(\ref{Mass_temp}) and (\ref{eq:S_bh}). The perturbative approximation is valid in the regime
\begin{equation}
    \alpha \ll r_+,
\end{equation}
and therefore all conclusions are restricted to this domain.

For two identical black holes with initial mass $M_i$, charge $Q_i$, and horizon radius $r_i$, conservation of charge implies
\begin{equation}
    Q_f=2Q_i,
\end{equation}
while the second law of black-hole thermodynamics requires
\begin{equation}
    S_f\ge S_1+S_2=2S_i.
\end{equation}
The equality corresponds to the minimum allowed entropy of the remnant and therefore determines the lower bound on the final horizon radius.

Substituting the perturbative entropy,
\begin{equation}
S(r_+,\alpha)
=
\pi r_+^2
+\pi\alpha r_+
+\frac{\alpha^2}{2}
\ln\left(\frac{r_+}{\ell_0}\right)
+\mathcal O(\alpha^3),
\end{equation}
into the equality
\begin{equation}
S_f=2S_i,
\end{equation}
and expanding the final horizon radius as
\begin{equation}
r_f=r_0+\alpha r_1+\alpha^2r_2+\mathcal O(\alpha^3) \ ,
\end{equation}
one finds, order by order in $\alpha$,
\begin{align}
r_0=\sqrt2\,r_i \quad,\quad r_1=\frac{2-\sqrt2}{2\sqrt2} \, \quad,\quad r_2=\frac{\dfrac1\pi\left(\dfrac12\ln\left(\dfrac{r_i}{\ell_0}\right)-\dfrac14\ln2\right)-\dfrac14}{2\sqrt2\,r_i} \ ,
\end{align}
where $\ell_0$ denotes an arbitrary reference length introduced to render the logarithm dimensionless. Consequently, the minimum allowed horizon radius\footnote{Complete derivation in Appendix~\ref{app:entropy_bound}.} of the remnant is
\begin{equation}
r_f^{\rm min}=\sqrt2\,r_i+\frac{2-\sqrt2}{2\sqrt2}\alpha+\frac{\alpha^2}{2\sqrt2\,r_i}\left[\frac1\pi\left(\frac12\ln\left(\frac{r_i}{\ell_0}\right)-\frac14\ln2\right)-\frac14\right]+\mathcal O(\alpha^3) \ .
\label{rfbound}
\end{equation}
Now using the same approximation simplifies the calculation of thermodynamic quantities, including the mass and entropy, and enables a direct comparison with RN-AdS black holes without NC corrections. Initial and final black hole masses to second order in \(\alpha\) take the following form
\begin{equation}
    M_{i,f}=\dfrac{l^2 Q^2_{i,f}+l^2r_{+_{i,f}}^2+r_{+_{i,f}}^4}{2l^2 r_{+_{i,f}}^4}+\dfrac{\alpha(l^2 r_{+,i,f}^2-l^2 Q^2_{i,f})}{4l^2 r_{+_{i,f}}^2}+\dfrac{\alpha^2(r_{+_{i,f}}^2+l^2)}{8l^2 r_{+_{i,f}}}~.
\end{equation}
Here, subscripts \(i\) and \(f\) denote initial and final black hole parameters. 

\begin{figure}[h!]
    \centering
    \includegraphics[width=0.6\textwidth]{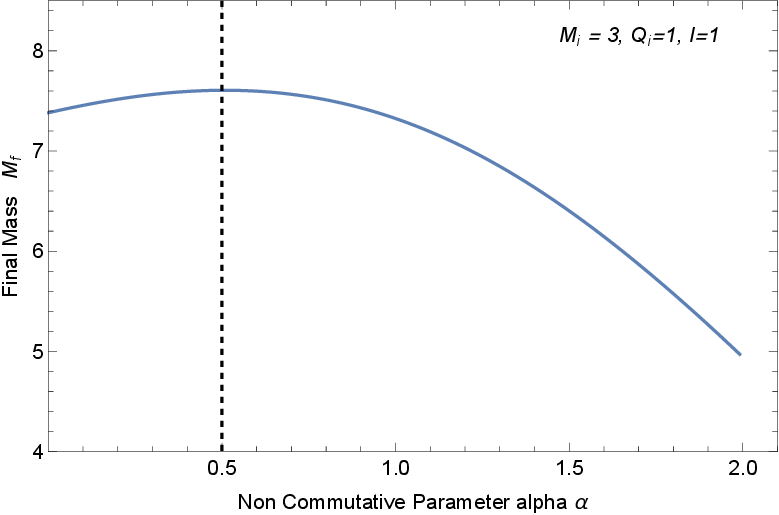}
    \caption{Final black-hole mass \(M_f\) as a function of the NC parameter \(\alpha\) for \(M_i = 3\), \(Q_i = 1\), and \(l = 1\).}
    \label{fig:F1}
\end{figure}

First, we plot the final mass \(M_f\) as a function of the NC parameter \(\alpha\) for an initial mass \(M_i = 3\), AdS length scale \(l = 1\), and charge \(Q_i = 1\), as shown in Figure~\ref{fig:F1}. This choice fixes the final charge on the black hole to \(Q_f=2\) and parameter \(M_f\) remains only as function of \(\alpha\). Although the full range of \(\alpha\) cannot be regarded as physically reliable because the perturbative expansion is valid only in the regime \(\alpha \ll r_{+_{i,f}}\), where \(r_{+_{i,f}}\) are the horizon radii of initial and final black holes, the small-\(\alpha\) behavior of the final-mass bound is noteworthy. The bound initially increases as the system deviates from the RN-AdS limit, reaches a maximum, and subsequently decreases. This indicates that the NC parameter has a significant effect on the post-merger bound on the final black-hole mass.

\begin{figure}[h!]
    \centering
    \includegraphics[width=0.6\textwidth]{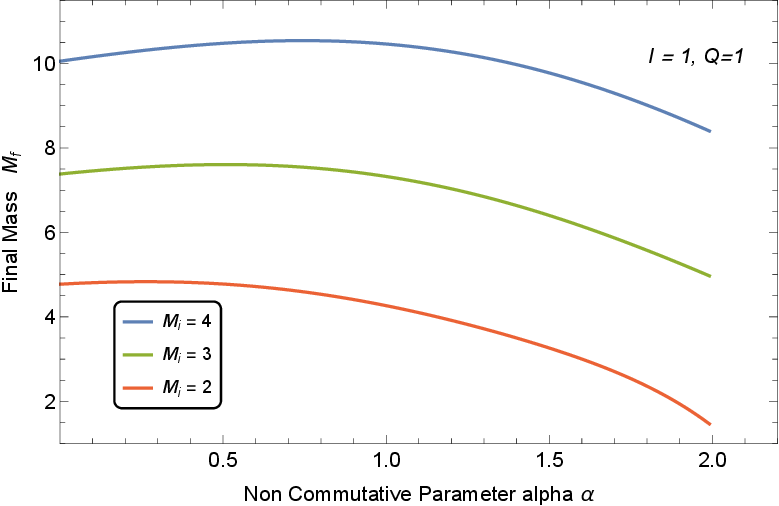}
    \caption{Final black-hole mass \(M_f\) as a function of the NC parameter \(\alpha\) for \(M_i = 2, 3, 4\), with \(Q_i = 1\) and \(l = 1\).}
    \label{fig:F2}
\end{figure}

\begin{figure}[h!]
    \centering
    \includegraphics[width=0.6\textwidth]{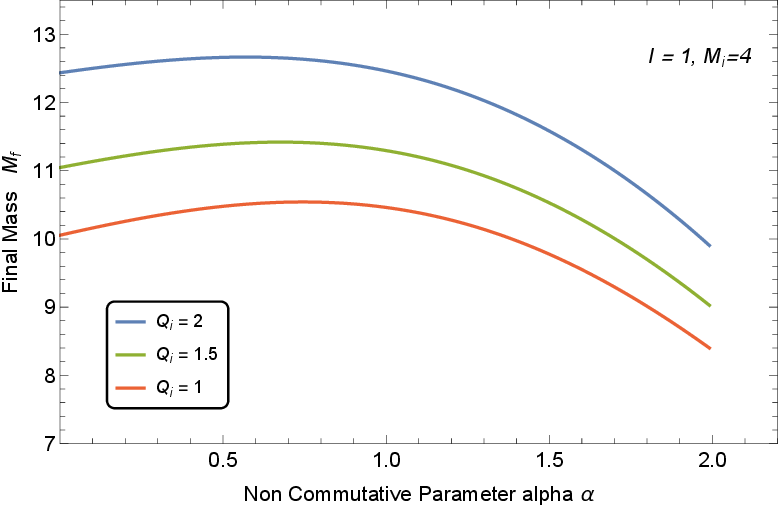}
    \caption{Final black-hole mass \(M_f\) as a function of the NC parameter \(\alpha\) for \(M_i = 4\), \(Q_i = 1, 1.5, 2\), and \(l = 1\).}
    \label{fig:F3}
\end{figure}

We next examine the dependence of the final-mass bound on the initial mass and charge, as shown in Figures \ref{fig:F2} and \ref{fig:F3}. The charge values are chosen such that the black holes remain away from the extremal limit. Figure \ref{fig:F2} shows that increasing the initial mass increases the final-mass bound. For each initial mass, the bound again exhibits a non-monotonic dependence on \(\alpha\), increasing at small \(\alpha\), reaching a maximum, and then decreasing. Figure \ref{fig:F3} demonstrates that, for a fixed initial mass, varying the charge also modifies the final-mass bound. In particular, increasing the charge leads to a higher bound over the range of \(\alpha\) considered. As before, only the small-\(\alpha\) region is within the perturbative regime, and conclusions drawn from the large-\(\alpha\) region of the plots should therefore be interpreted with caution.\\

In conclusion, the NC correction parameter significantly modifies the lower bound on the final black-hole mass following a merger. These results suggest that NC effects can alter black-hole thermodynamics and may influence the amount of energy released through gravitational radiation. The modification of the minimum allowed final mass directly translates into a corresponding change in the maximum amount of gravitational-wave energy that can be emitted during the merger. Although the parameter $\alpha$ is expected to be extremely small, these results provide an analytical framework for understanding how quantum-geometric corrections may influence black-hole merger thermodynamics and the associated energetics.

\section{Summary and Discussions}\label{Sec:Summary and Discussions}
In this work, we have explored the charged–AdS black holes in NC spacetime with Lorentzian-smeared distributions. Subsequently, we investigated the thermodynamical properties of these black holes within the framework of NC geometry. Specifically, we analyzed the thermal stability and the critical behavior. To this end, we computed the thermodynamic quantities needed to capture the system's physical features. Our analysis was carried out within a perturbative framework. This approach is particularly well-motivated in the context of NC geometry, where the deformation parameter introduces a fundamental length scale expected to be small compared to the black hole's characteristic size. Treating noncommutativity as a perturbation, therefore, allows one to capture its leading-order effects on black hole thermodynamics without sacrificing analytic control. Such a perturbative expansion not only provides clear physical insight but also offers a systematic way to smoothly connect to the commutative limit, thereby highlighting how even small NC corrections can imprint themselves on the stability structure and phase behavior of AdS black holes.

The topological classification of the boundary conformal field theory dual to the NC RN-AdS black hole yields a clear and robust physical distinction. The key outcome is the topological transition from $W=+1$ for the commutative ($\alpha=0$) boundary theory to $W=0$ for its NC ($\alpha>0$) counterpart. This change is not merely quantitative; it reflects a fundamental shift in the nature of the gravitational dual: $W=+1$ characterizes CFTs dual to classical, singular spacetimes, while $W=0$ signals a dual bulk geometry regularized by a minimal length scale. The mechanism driving this transition is universal---for any $\alpha>0$, the small-horizon asymptotic behavior of the CFT temperature $\tilde{T}(r_+)$ flips from $-\infty$ to $+\infty$. This creates a novel, ultra-small black hole (USBH) branch with negative winding ($ w =- 1$), thereby altering the global topological sum. Crucially, this result is independent of the specific values of $Q$, $\ell$, $c$, and the boundary volume $\mathcal{V}$, provided the physical conditions $Q>0$, $\ell>0$, and $\alpha>0$ are maintained. It persists across both subcritical and supercritical pressures. Therefore, the topological invariant $W$ serves as a sharp, non-perturbative diagnostic in the holographic dictionary: it can distinguish, from the boundary CFT data alone, whether the dual gravitational description is classically singular or incorporates UV-completing quantum-gravitational features. This work demonstrates the power of topological thermodynamics to extract deep structural information about quantum gravity from boundary observables and show a match between bulk and boundary topological charges.

Finally, we observe that the NC correction parameter significantly modifies the lower bound on the final black-hole mass following a merger. These results suggest that NC effects can alter black-hole thermodynamics and may influence the amount of energy released through gravitational radiation.




\section*{Acknowledgments}

Ankit Anand is financially supported by the institute's postdoctoral fellowship at IIT Kanpur. Aditya Singh acknowledges support from the Council of Scientific and Industrial Research-Human Resource Development Group (CSIR-HRDG), under the Postdoctoral Research Associate Fellowship, funded through Project No. 03WS(003)/2023-24/EMR-II/ASPIRE.


\appendix


\appendix
\section{Perturbative derivation of the minimum horizon radius}
\label{app:entropy_bound}

In this appendix, we derive the perturbative expression for the minimum horizon radius of the remnant black hole resulting from the merger of two identical charged black holes. The derivation is based on the second law of black-hole thermodynamics and is valid up to second order in the NC parameter $\alpha$.

For two identical initial black holes, the second law implies
\begin{equation}
S_f\ge S_1+S_2=2S_i.
\end{equation}
The equality corresponds to the minimum allowed entropy of the remnant and therefore determines the lower bound on the final horizon radius. Using the perturbative entropy formula,
\begin{equation}
S(r_+,\alpha)=
\pi r_+^2
+\pi\alpha r_+
+\frac{\alpha^2}{2}
\ln\left(\frac{r_+}{\ell_0}\right)
+\mathcal{O}(\alpha^3),
\end{equation}
where $\ell_0$ is an arbitrary reference length introduced to render the logarithm dimensionless, the entropy equality becomes
\begin{equation}
\pi r_f^2
+\pi\alpha r_f
+\frac{\alpha^2}{2}
\ln\left(\frac{r_f}{\ell_0}\right)
=
2\pi r_i^2
+
2\pi\alpha r_i
+
\alpha^2
\ln\left(\frac{r_i}{\ell_0}\right).
\label{entropyeq}
\end{equation}

Since the NC parameter is assumed to satisfy $\alpha\ll r_i$, we seek a perturbative solution of the form
\begin{equation}
r_f
=
r_0
+\alpha r_1
+\alpha^2r_2
+\mathcal{O}(\alpha^3).
\label{rfexpand}
\end{equation}

The square of the horizon radius is readily expanded as
\begin{align}
r_f^2 =
(r_0+\alpha r_1+\alpha^2r_2)^2 =
r_0^2
+
2\alpha r_0r_1
+
\alpha^2
\left(
r_1^2+2r_0r_2
\right)
+\mathcal{O}(\alpha^3) \ .
\label{rf2}
\end{align}

Similarly,
\begin{align}
\ln r_f =
\ln\left(
r_0+\alpha r_1+\alpha^2r_2
\right)=
\ln r_0
+
\ln\left(
1+\frac{\alpha r_1+\alpha^2r_2}{r_0}
\right).
\end{align}
Using the Taylor expansion
\begin{equation}
\ln(1+x)
=
x-\frac{x^2}{2}
+\mathcal{O}(x^3),
\end{equation}
one finds
\begin{equation}
\ln r_f
=
\ln r_0
+
\frac{\alpha r_1}{r_0}
+
\alpha^2
\left(
\frac{r_2}{r_0}
-
\frac{r_1^2}{2r_0^2}
\right)
+\mathcal{O}(\alpha^3).
\end{equation}
Since the logarithm is multiplied by $\alpha^2$ in Eq.~(\ref{entropyeq}), only its leading contribution is required,
\begin{equation}
\frac{\alpha^2}{2}
\ln\left(\frac{r_f}{\ell_0}\right)
=
\frac{\alpha^2}{2}
\ln\left(\frac{r_0}{\ell_0}\right)
+\mathcal{O}(\alpha^3).
\end{equation}

Substituting these expansions into Eq.~(\ref{entropyeq}) and matching equal powers of $\alpha$ yields a sequence of algebraic equations.

At zeroth order,
\begin{equation}\label{first_order}
r_0^2 = 2r_i^2,\qquad \implies   r_0
=
\sqrt2\,r_i \ .
\end{equation}

At first order,
\begin{equation}
2r_0r_1+r_0 = 2r_i \qquad \implies r_1
=
\frac{2r_i-r_0}{2r_0} \ .
\end{equation}
Using the zeroth-order solution as in Eq.~\eqref{first_order}, we have 
\begin{equation}
r_1 = \frac{2-\sqrt2}{2\sqrt2} \ .
\end{equation}

At second order,
\begin{equation}
\pi
\left(
r_1^2
+
2r_0r_2
+
r_1
\right)
+
\frac12
\ln\left(\frac{r_0}{\ell_0}\right)
=
\ln\left(\frac{r_i}{\ell_0}\right),
\end{equation}
which gives
\begin{equation}
r_2
=
\frac{
\dfrac1\pi
\left[
\ln\left(\dfrac{r_i}{\ell_0}\right)
-
\dfrac12
\ln\left(\dfrac{r_0}{\ell_0}\right)
\right]
-r_1-r_1^2
}
{2r_0}.
\end{equation}

Using Eq.~\eqref{first_order} together with
\begin{equation}
\ln\left(\frac{r_0}{\ell_0}\right)
=
\ln\left(\frac{r_i}{\ell_0}\right)
+\frac12\ln2,
\end{equation}
one obtains
\begin{equation}
r_2
=
\frac{
\dfrac1\pi
\left[
\frac12
\ln\left(\dfrac{r_i}{\ell_0}\right)
-
\frac14\ln2
\right]
-\dfrac14
}
{2\sqrt2\,r_i} \ .
\end{equation}

Collecting all contributions, the perturbative lower bound on the horizon radius of the remnant becomes
\begin{equation}
r_f^{\rm min}
=
\sqrt2\,r_i
+
\frac{2-\sqrt2}{2\sqrt2}\alpha
+
\frac{\alpha^2}{2\sqrt2\,r_i}
\left[
\frac1\pi
\left(
\frac12
\ln\left(\frac{r_i}{\ell_0}\right)
-
\frac14\ln2
\right)
-\frac14
\right]
+\mathcal{O}(\alpha^3) \ .
\end{equation}

This perturbative expression is as in Eq.~\eqref{rfbound}.



\bibliographystyle{utphys.bst}
\bibliography{ref}



\end{document}